\documentclass[aps,pre,twocolumn,showpacs,superscriptaddress,groupedaddress]{revtex4}
\usepackage{graphicx}  % needed for figures
\usepackage{dcolumn}   % needed for some tables
\usepackage{bm}        % for math
\usepackage{amssymb}   % for math

\begin{document}

\title{Perfluorooctanoic acid rigidifies a model lipid membrane}
\author{B. Br{\"u}ning}
\author{B.~Farago}

\date{\today}

\begin{abstract}
We report a combined dynamic light scattering (DLS) and neutron spin-echo (NSE) study on vesicles composed of the phospholipid 1,2-dimyristoyl-sn-glycero-3-phosphatidylcholine (DMPC) under the influence of varying amounts of perfluorooctanoic acid (PFOA). We study local lipid bilayer undulations using NSE on time scales up to 200 ns. Similar to the effect evoked by cholesterol, we attribute the observed lipid bilayer stiffening to a condensing effect of the perfluorinated compound on the membrane.
\end{abstract}

\maketitle

Perfluorinated compounds (PFCs) are fully fluorinated fatty acid analogues commonly used in a wide range of applications, such as the production of fire-extinguishing foams, anticorrosion agents, lubricants or cosmetics \cite{Kissa2001,Riess2009}. As a consequence of their chemical stability, these compounds exhibit an environmental stability and are transmitted into the mammal food chain \cite{Ropers2009}. Due to their tendency to bioaccumulate, the compounds affect properties of cell membranes, causing developmental and reproductive disorders \cite{Riess2009,Matyszewska2008a}. It is of particular interest to understand the effect of perfluorinated compounds not only on cellular membranes \cite{Nakahara2012,Krafft2001}, but also on their biomimetic counterparts \cite{Zaggia2012}. In mammal organisms, vesicular membranes often serve as natural carriers. It is assumed that functional properties of a membrane depend likewise on its composition-dependent structure and dynamics \cite{Bruening2010,Rheinstaedter2008}. In order to gain insight into membrane function, specific material properties, such as e. g. the bilayer bending rigidity $\kappa$ can be aimfully influenced. Several studies have addressed the insertion of perfluorinated compounds into binary model membranes. Oriented mono- and bilayers were investigated by Matyszewska et al. using methods including surface pressure and potential measurements, infrared spectroscopy (IR), nuclear magnetic resonance (NMR) techniques and molecular dynamics (MD) simulations \cite{Matyszewska2008a,Matyszewska2007,Matyszewska2008}. While the phospholipid headgroup tilt against the bilayer normal decreases with rising amounts of inserted PFOA, the lipid acyl chain order increases. Lateral molecule diffusion in the membrane plane changes non-uniformly as the ratio of the components in the  binary mixture is varied \cite{Matyszewska2007}. Lehmler et al. study liposomes containing binary mixtures of varying phospholipids and perfluorinated surfactants \cite{Lehmler2006,Xie2010,Xie2007,Xie2010a}. They use fluorescence spectroscopy and differential scanning calorimetry measurements to study the partitioning of surfactants into the phospholipid bilayer and find that it is independ of the lipid acyl chain length. While the phase behaviour is largely independent of the type of phospholipid, PFOA itself is found to partition more readily into lipid bilayers in their fluid phase \cite{Xie2010}. Several studies have taken advantage of a combination of DLS and long-wavelength NSE for the investigation of the local bilayer undulation dynamics in unilamellar lipid vesicles (ULV's) \cite{Arriaga2009a,Arriaga2009,Arriaga2010,Bruening2013,Mell2013,Hirai2013}. Here, we cover a window of more than 200~ns.
The membrane dynamics was investigated by NSE using wavelengths of $\lambda=10$~\AA $ $ and 17~\AA. We  highlight complementary dynamic regimes by adapting the time range of the correlation function decay. We use our DLS-results for a quantitative separation of two dynamical processes in the 17 \AA $ $ data.

The phospholipid DMPC was purchased from Avanti (Alabaster, AL), the perfluorinated surfactant PFOA from Sigma-Aldrich (Steinheim, Germany). Components were dissolved in chloroform in the desired molar proportions. The solvent was evaporated and the dry lipids were hydrated with heavy water (D$_2$O) at 10~mg/ml. D$_2$O was used both for dynamic light scattering (DLS) and neutron spin-echo (NSE). For each specific composition, the PH was compensated through the addition of Na$_2$CO$_3$-base. To obtain unilamellar vesicles (ULV's), the suspension was passed ten times through a polycarbonate filter with 500~\AA $ $ pore diameter using a LiposoFast Basic Extruder (A\-ves\-tin, Ca\-na\-da). For the NSE experiment, samples were poured into 1~mm thick quarz cuvettes with a quadratic cross section of 35~mm by 35~mm  (Hellma, M\"ullheim, Germany).

\begin{figure}
\resizebox{1.0\columnwidth}{!}{%
 \includegraphics{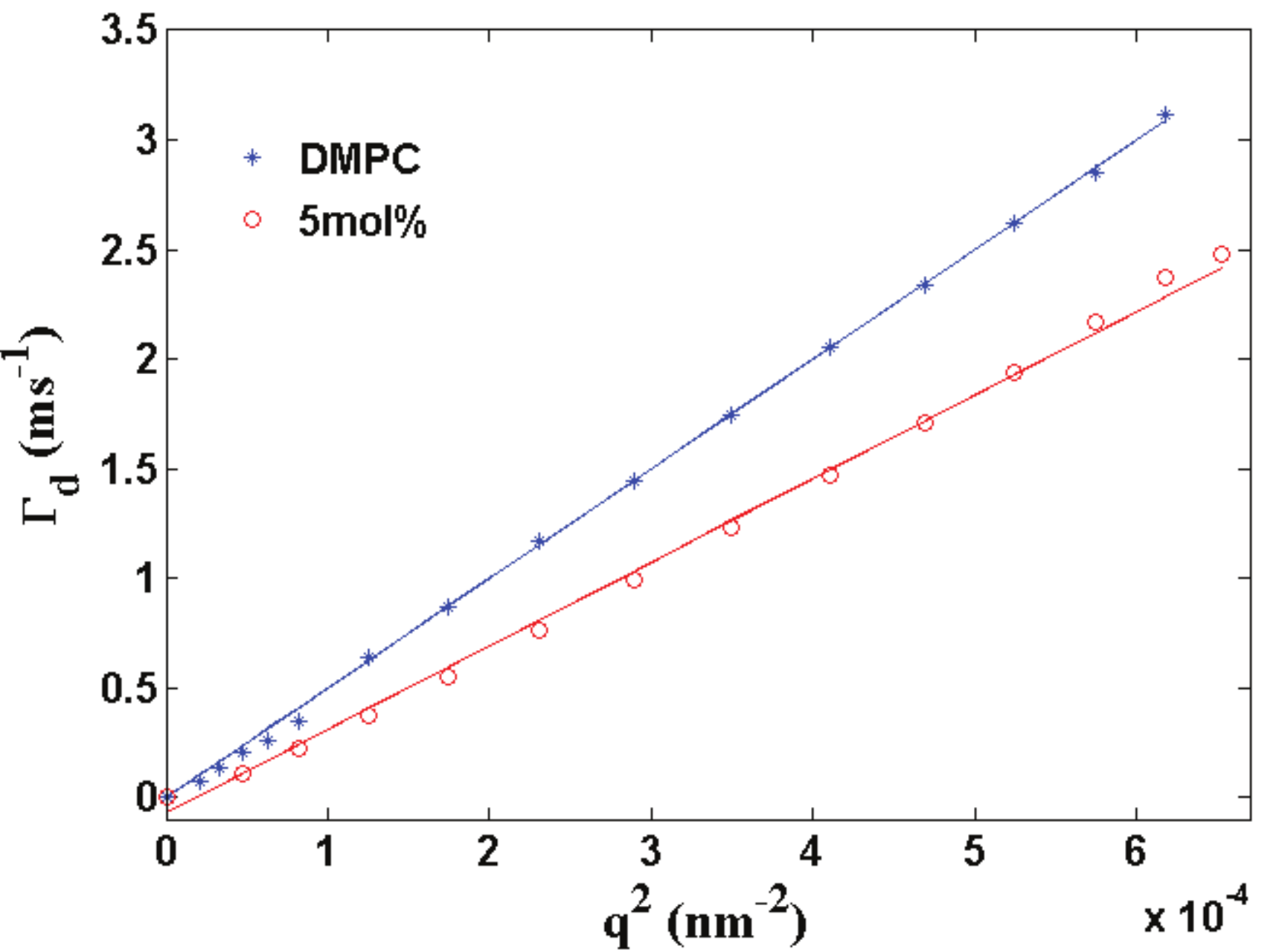} } \caption{\label{fig:DMPCPFOA_DLSangles} (color online).
Relaxation rate $\Gamma_d$ \emph{vs.} $q^2$, for vesicles composed of DMPC and DMPC/PFOA (5mol\%). Linear fits yield the vesicle center-of-mass diffusion constant D.}
\end{figure}

For dynamic light scattering (DLS), an ALV goniometer with a 35~mW He-Ne laser operating at a wavelength of 632.8~nm was used with an ALV/High QE APD detector and an ALV-6010/ 160 external multiple $\tau$ digital correlator unit. The vesicle center-of-mass diffusion can be described by a correlation function \begin{math} g_1(t) = \exp(-D \cdot q^2  t) \end{math}, which is derived from the measured intensity correlation function $g_2(t)$ through the Siegert relation \begin{math} g_1(t)=\sqrt{g_2(t)-1}  \end{math}. The center-of-mass diffusion coefficient $D$ and the hydrodynamic vesicle radius $R_H$ are linked according to the Stokes-Einstein equation: \begin{math} R_H=\frac{k_B T}{6 \pi \cdot \eta(T) \cdot D} \end{math}. $k_B$ denotes the Boltzmann constant, $T$ the absolute temperature and $\eta(T)$ the temperature-dependent solvent viscosity (for D$_2$O at $30^\circ$~C,  $\eta(T)=0.973\cdot10^{-3}$~Pa~s). Long term measurements indicate that the averaged vesicle radii $R_H$ remained constant for at least a week. The vesicles for the combined DLS and NSE experiment were freshly prepared and measured within one day. Angle-dependent measurements were taken between 20$^\circ$ and 150$^\circ$ in steps of 10$^\circ$, the obtained relaxation rates $\Gamma_d$  \emph{vs.} $q^2$ are shown in fig.~\ref{fig:DMPCPFOA_DLSangles}. Linear fits indicate a purely Fickian diffusion. The linear slope of these curves corresponds to the vesicle center-of-mass diffusion constant $D$.

The neutron spin-echo (NSE) experiment was performed at the cold spectrometer IN15 at the Institut Laue Langevin (ILL, Grenoble, France). Due to its fine angular resolution in the small-angle regime (small $q$), the instrument is well suited to probe mesoscopic length scales. An introduction into the NSE method is given in~\cite{Mezei1980}. A measurement yields a momentum transfer and time-resolved intermediate scattering function $S(q,t)$, in which the Fourier time $t$ changes proportionally to the wavelength $\lambda^3$  and the applied magnetic field integral, following  \begin{math} t \propto \lambda^3 \int |B| dl\label{eqn:Def_Fouriertime} \end{math}. Error bars in the measured data correspond to $\pm \sigma$ statistical error calculated from the counting statistics and are transferred to our least squares fits. The $q$-range probed by NSE lies in the range of the inverse length scales of local bilayer interface undulations. This allows a data interpretation on the basis of models including a unique $q$-dependence of the measured relaxation rates $\Gamma(q)$. On IN15, incident neutron wavelengths between 6 and 25~\AA$ $ with a wavelength spread of 15\% are available. At a distance of 4.6~m from the sample, a He-3/CF$_4$ multidetector is located with $32 \cdot 32$~pixels of 1~cm$^2$ each. The instrumental resolution is determined using graphoil as a purely coherent elastic scatterer and accounted for by division of sample scattering through resolution signal \cite{Farago2009}.

\begin{figure}
\resizebox{1.0\columnwidth}{!}{%
 \includegraphics{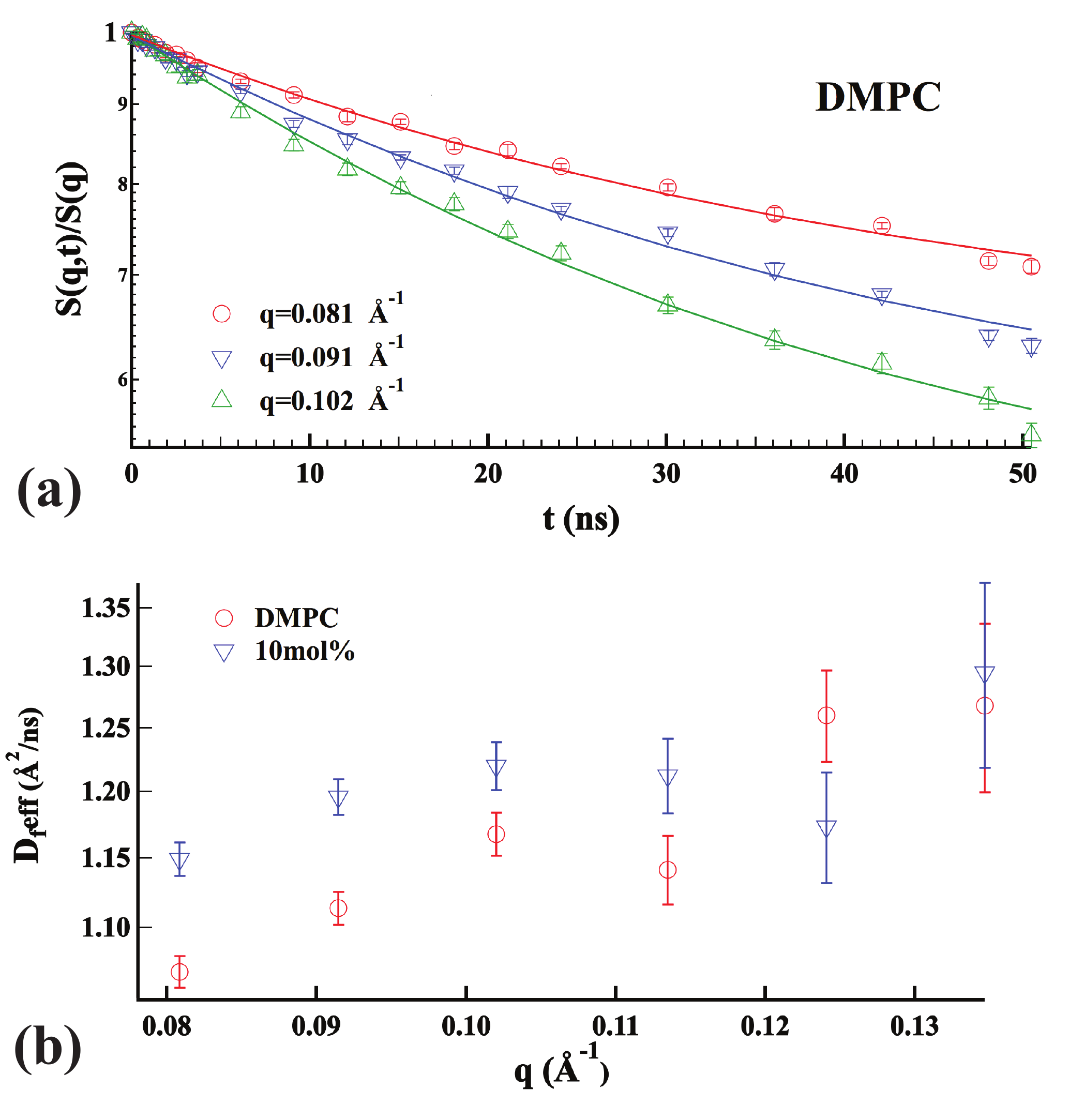} } \caption{\label{fig:SurfSefit_examps_Dqh2} (color online).
(a) Normalized intermediate scattering function~$S(q,t)/S(q)$ for DMPC (30$^{\circ}$C) at $\lambda=10$~\AA.
Single-exponential fits yield the relaxation rates~$\Gamma_f(q)$; (b) the effective diffusion constant $D{_f}{eff}(q)=\Gamma_f(q)/q^2$ exhibits a $q$-dependence, indicating further dynamics. }
\end{figure}

Merely simple diffusion mechanisms were assumed for the short wavelength data $S(q,t)\propto\exp(-\Gamma_{f} t)$, with a relaxation rate $\Gamma_{f}=D_f \cdot q^2$. An example is shown in fig.~\ref{fig:SurfSefit_examps_Dqh2}~(a) for DMPC at 30$^{\circ}$C, within the $q$-range in which the most pronounced changes in the dynamics occur. Fig.~\ref{fig:SurfSefit_examps_Dqh2}~(b) then shows an effective diffusion constant $D_f eff(q)$ derived as $\Gamma_f(q)/q^2$ for DMPC and DMPC/PFOA (10mol\%). In both cases, this plot is not $q$-independent, as would be the case if only Fickian diffusion occured in the probed time range. Unfortunately, the different dynamic contributions cannot be unambigiously  separated, due to a lack of a full polarization decay at the largest measured $q$ shown in~(a). The dynamics are not fully covered within the probed window, necessitating our choice of a higher wavelength to extend the Fourier-time range. Here, a smaller $q$-range is covered with higher resolution, thus two complementary data sets are obtained. Regarding the origin of the quadratic mode $\Gamma_f(q^2)$, several possibilities exist: In their seminal theoretical works, Evans and Yeung, as well as Seifert and Langer have already predicted modes, which relate bending and local density changes between the two monolayers and taken into account resulting intermonolayer friction \cite{Evans1994,Seifert1993}. The modes follow quadratic relations $\Gamma(q^2)$, and have been discussed in recent experimental works by Arriaga et al. \cite{Arriaga2010}. On the other hand, the contribution lies close to the one of the vesicle center-of-mass diffusion found by dynamic light scattering. Following, we discuss vesicle center-of-mass translations and bilayer undulations as separable dynamic contributions in the long wavelength spin-echo data.

Curvature undulations of elastic membranes are commonly described by the well-known Helfrich Hamiltonian \cite{Helfrich1973}. Based on this continuum mechanical approach Milner and Safran further describe the fluctuation dynamics of microemulsion drop\-lets and vesicles \cite{Milner1987}. In their theory, the normal bending modes of the flexible interface are coupled to the viscous friction exerted by the suspending medium according to a single exponential decay $\exp(-\Gamma_{MS}t)$ with a relaxation rate $\Gamma_{MS}=\frac{\kappa}{4 \eta}q^3$, where $\eta$ is the effective viscosity of the solvent medium and $\kappa$ the bilayer bending rigidity. Faster relaxations are assigned to stiffer membranes. While suited to describe soft interfaces with bending rigidities on the order of $k_B T$, such as micro-emulsion droplets and sponge phases,  the expression fails to accurately account for dynamics of model lipid membranes with bending rigidities of several $k_B T$. Zilman and Granek describe curvature shape fluctuations of freely suspended phospholipid bilayers \cite{Zilman1996,Zilman2002}.  Their model takes into account a coupling of the bending modes and local diffusion processes: in a rigid membrane with a bilayer bending rigidity of ($\kappa \gg k_B T$), less free volume can be explored by single molecules; this means that a relaxation rate for a coupled process of undulation and local curvature will increase, whereas the average amplitude of the modes will decrease. The anomalous subdiffusive relaxation of the bending motions is described by a stretched exponential decay with a stretching exponent of $\beta=2/3$ (eq.~\ref{eq: ZGdecay}):

\begin{eqnarray}
S(q,t)&\propto& \exp(-\Gamma_u(q) \cdot t)^\beta \nonumber\\ \mbox{with}
\qquad \Gamma_u(q)&=&0.025\gamma_q\left(\frac{k_B
T}{\kappa}\right)^{1/2}\cdot \left(\frac{k_B T}{\eta(T)} \right)q^3
\label{eq: ZGdecay}
\end{eqnarray}

\noindent The relaxation rate $\Gamma_u(q)$ includes the tem\-pe\-ra\-ture-de\-pen\-dent solvent viscosity $\eta(T)$. Further, $\gamma_q$ is a weak monotonous function of the bending rigidity $\kappa$ according to \begin{math} \gamma_q=1-\frac{3}{4 \pi}\left(\frac{k_B T}{\kappa} \right) \cdot \ln(qh)\end{math}, where $h$ is the membrane thickness with \begin{math} q \cdot h \approx 1 \end{math}. When $\kappa$ lies on the order of several $k_B T$, $\gamma_q$ can be approximated to unity. Using atomistic and coarse-grained molecular dynamics (MD) simulations, Brandt and Edholm describe the nanometer length scale fluctuation decay in fluid biomembranes by a stretched exponential \cite{Brandt2010}. We discuss our experimental data on the basis of this approach, augmented by the occurrence of underlying vesicle center-of-mass diffusion (eq.~\ref{eq:combinedfit}). $A$ is a normalization parameter close to one, the mass diffusion relaxation rate $\Gamma_d=D\cdot q^2$ is fixed, the relaxation rate of bilayer undulations  $\Gamma_u$ is a free parameter, and the stretched exponential is held to  $\beta=0.66$, following the Zilman-Granek approach. With respect to previous works, such as \cite{Arriaga2009a,Arriaga2009}, this seems to be a more general approach. Comparing the outcome to the results obtained on the basis of the previous description of S(q, t)/S(q), we find no significant differences within  experimental precision.

\begin{equation}
S(q,t)/S(q,0)=A \cdot \exp(-\Gamma_{d} t)\cdot
\exp(-\Gamma_u t)^{\beta} \label{eq:combinedfit}
\end{equation}

\begin{figure}
\resizebox{1.0\columnwidth}{!}{%
 \includegraphics{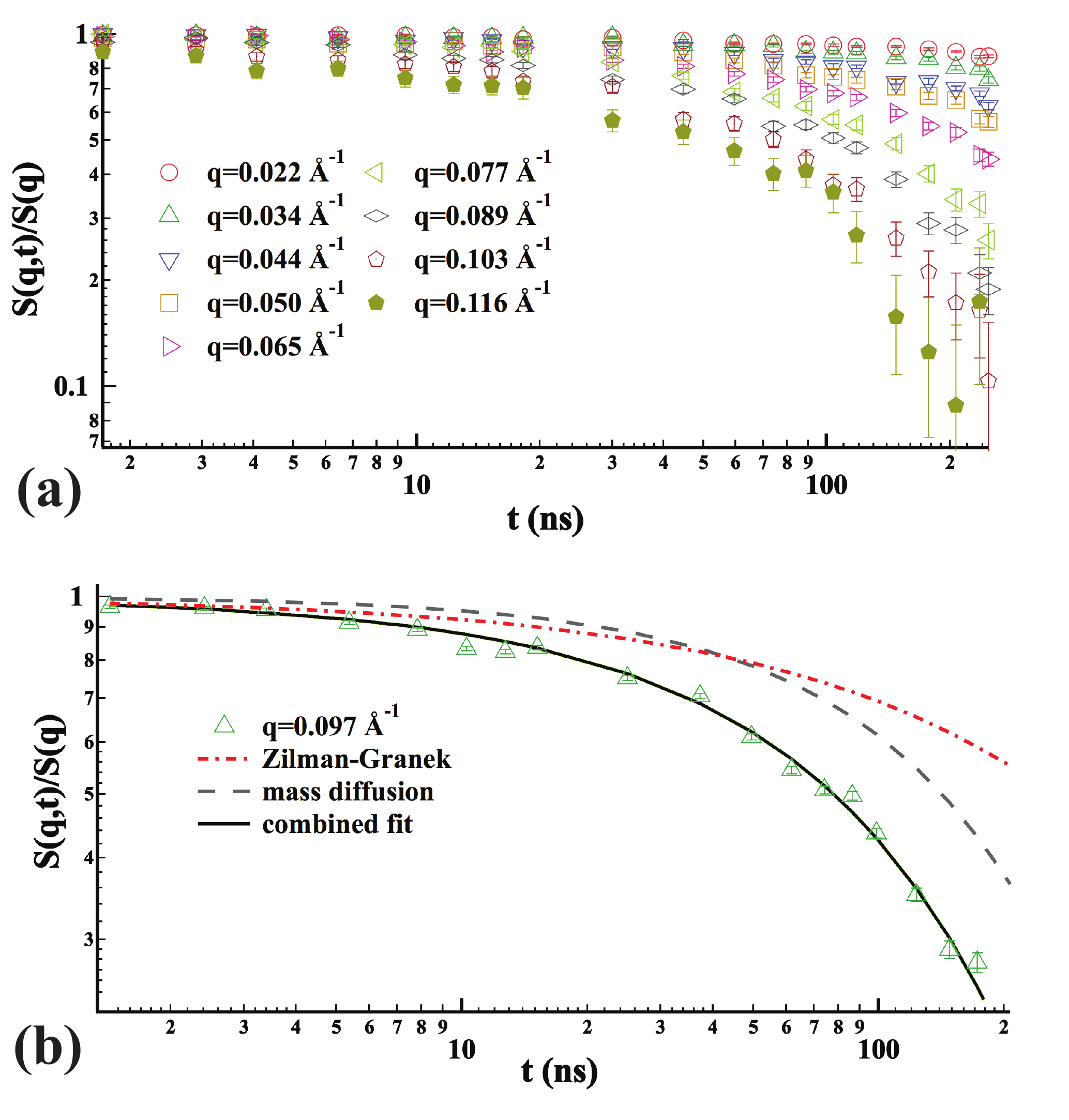} } \caption{\label{fig:DMPCstandardfit_16A} (color online).
 Normalized intermediate scattering functions~$S(q,t)/S(q)$ for DMPC (30$^\circ$C): (a) $q$-range covered at $\lambda=17$~\AA; (b) combined fit  according to eq.~\ref{eq:combinedfit}, single contributions are indicated by dotted lines.}
\end{figure}

\begin{figure}
\resizebox{1.0\columnwidth}{!}{%
 \includegraphics{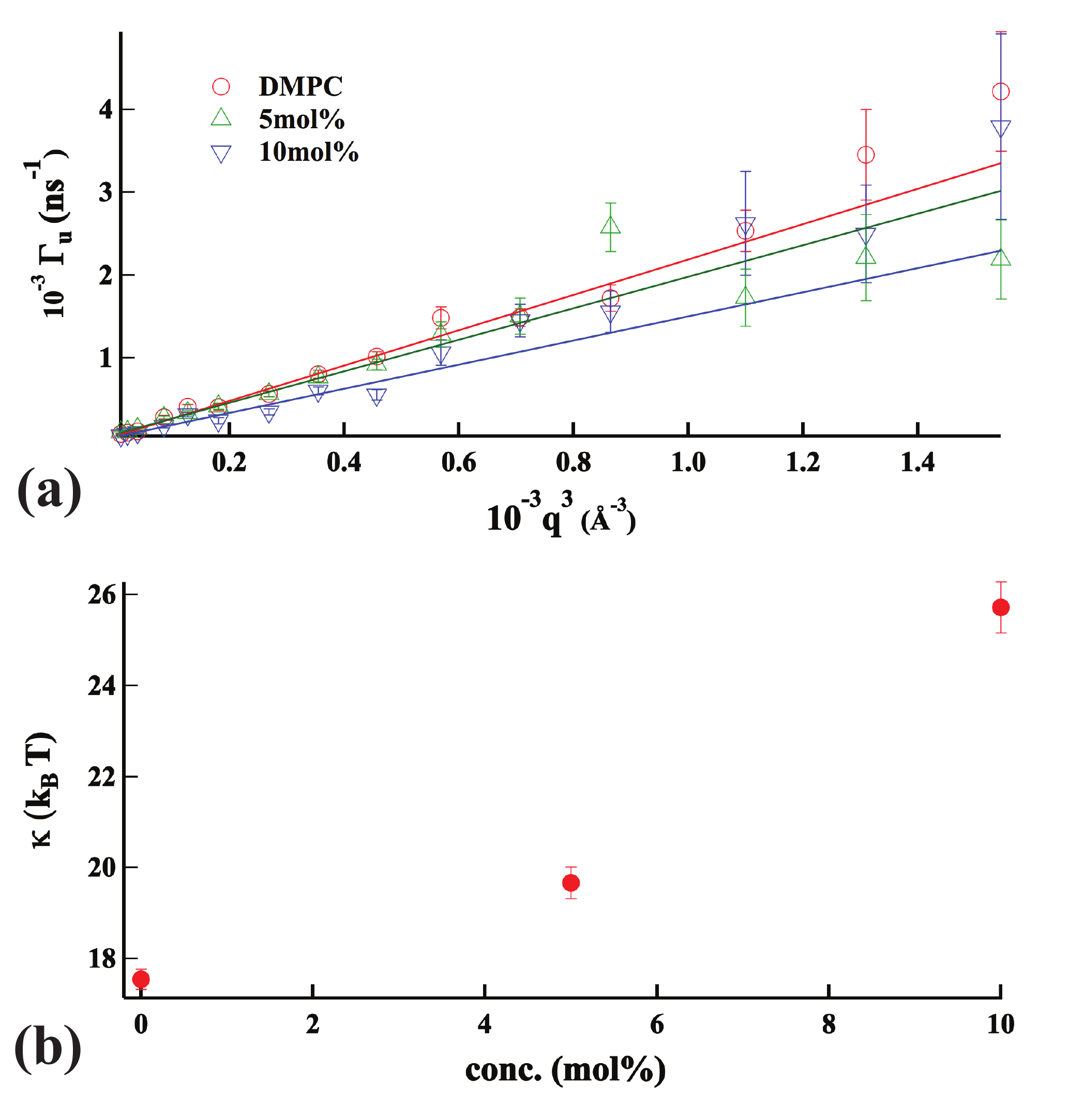} } \caption{\label{fig:SurfactantSpinEchoGammau_ZGDLS} (color online).
Bilayer undulations: (a) Relaxation rate $\Gamma_u$ \emph{vs.} $q^3$, for DMPC and DMPC/PFOA. Linear fits yield the concentration dependent bilayer bending rigidity $\kappa$ shown in~(b).}
\end{figure}

The normalized intermediate scattering function of the DMPC standard $S(q,t)/S(q)$ is shown in fig.~\ref{fig:DMPCstandardfit_16A}~(a) for varying $q$. Fourier times extend up to 200~ns, and the normalized polarization nearly fully decays. This means, the dynamics are captured within the spectrometer window, enabling more precise treatment than before. Over the whole $q$-range, fits can be improved taking into account dynamic contributions just outside the instrumental window in the $\mu$s-regime. Therefore, we add a vesicle center-of-mass diffusion contribution through the diffusion constant~$D$ obtained from DLS measurements (eq.~\ref{eq:combinedfit}). In fig.~\ref{fig:DMPCstandardfit_16A}~(b), both vesicle center-of-mass diffusion, as well as bilayer undulation contributions are shown for one exemplary $q$-value ($q=0.097$~\AA).  The double logarithmic scale shows that the decay curvature is well matched by the combined fit. At momentum transfers of $q=0.071$~\AA$ $ and above, a deviation from the single-exponential behaviour is observed, starting at Fourier times between 50 and 100~ns, cf.~(a). The undulation relaxation rate $\Gamma_u(q)$ can now be further analyzed. For our  DMPC standard in its fluid phase ($30^\circ$C), the result of $\kappa=17.68\pm0.15$ $k_B T$ derived from linear regression of $\Gamma_u(q^3)$ is well in accordance with literature values \cite{Mecke2003,Rheinstadter2006,Safinya1989}. The inherent experimental error for $\kappa$ is estimated to lie on the order of $k_B T$. The undulation relaxation rate $\Gamma_u(q^3)$ obtained from these fits are shown in fig.~\ref{fig:SurfactantSpinEchoGammau_ZGDLS}. Insertion of rising amounts of PFOA evokes a decrease in the linear slope of $\Gamma_u(q^3)$. The resulting increase in the bilayer bending rigidities $\kappa$ is shown in~(b).

%%%%%%%%%%%%%%%%%%%%%%%%%%%%%%%%%%%%%%%%%%%%%%%%%%%%%%%%%%%%%%%%%%%%%%%%%%%%%%
%%%%%%%%%%%%%%%%%%%%%%% Discussion%%%%%%%%%%%%%%%%%%%%%%%%%%%%%%%%%%%%%%%%%%%
%%%%%%%%%%%%%%%%%%%%%%%%%%%%%%%%%%%%%%%%%%%%%%%%%%%%%%%%%%%%%%%%%%%%%%%%%%%%%%
%\section{Summary \& Conclusions}\label{summary}

Comparing unilamellar vesicles composed of DMPC and DMPC/PFOA mixtures, we have investigated the effect of the perfluorinated compound on local bilayer undulations and on the bilayer bending rigidity $\kappa$. Effects on vesicle self-diffusion were also investigated. The NSE data can be meaningfully analyzed, assuming a combination of two separable contributions within the dynamic window up to 200~ns, namely vesicle center-of-mass diffusion and local bilayer undulations. The latter was described on the basis of the well-known Zilman-Granek approach for free film fluctuations. Our data supports a view that lipid bilayer undulation dynamics and corresponding bending rigidities $\kappa$ can be tuned by directly inducing changes at the lipid acyl chains: the perfluorinated compound PFOA, which partitions into the membrane \cite{Lehmler2006,Xie2010,Xie2007,Xie2010a}, reduces the free volume in the membrane plane, since the $CF_2$-segments of the surfactant tail occupy more space than the neighbouring lipid chain $CH_2$-segments. Consequently, an increase in the bilayer bending rigidity $\kappa$ is observed. In a previous work, we have also investigated binary lipid mixtures containing DMPC and the monounsaturated phospholipid DOPC \cite{Bruening2013}. Here, a significant decrease in $\kappa$ was observed, the higher the amount of the latter was. We suggested, this decrease might be explained by a mesoscopic lateral phase segregation leading to domain structure fluctuations and a corresponding softening of the membrane \cite{Bruening2009,Heimburg2007}. In the case of lipid/surfactant mixtures on the other hand, a homogenous lateral distribution of the two components is likely, similar to the one in binary cholesterol mixtures. The stiffening of a DMPC bilayer through cholesterol insertion has long since been associated with the condensing effect of the sterol \cite{Huang1999}. Nakahara et al. have suggested similar effects might occur for partially fluorinated alcohols in DPPC and DPPG membranes \cite{Nakahara2011}. In our case, the PFOA-induced stiffening of the DMPC bilayer might be explained as follows: The ionic surfactant head is more hydrophillic and the perfluorinated tail more hydrophobic than the zwitterionic phospholipid  headgroup and its acyl chains. Repulsion between neighboring surfactant heads as well as the bulkiness of the surfactant CF$_2$-segments in comparison to their CH$_2$ lipid acyl chain counterparts make an alternating in-plane molecule distribution of lipid and surfactant likely. The rotation of lipid acyl chain segments around the bilayer normal axis is then restricted. Thus, by PFOA-inclusion, the in-plane area density of lipid acyl chain and surfactant tail segments is then increased, causing lipid bilayer stiffening. Such a bilayer stiffening may well be the origin of the compounds harmful effects on cellular membranes, as it is likely to influence fusiogenic properties of lipid vesicles. Therefore, it would be interesting to study how the compounds influence the formation of stalk intermediates as precursor states for the fusion of lipid vesicles. This could further elucidate, how reproductive and developmental processes are inhibited in mammal organisms after perfluorinated compound ingestion.

%\section{Acknowledgements}\label{acknowledgements}

We are greatful to Ralf K{\"o}hler and Ralf Stehle for experimental support. Roland Steitz is thanked for helpful comments on the manuscript. We thank HZB for financial support and for the allocation of beam time on IN15, as well as ILL for technical support.

\bibliography{SurfactantSpinEchoPRE}

\end{document}